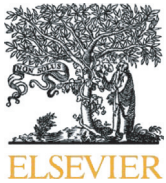
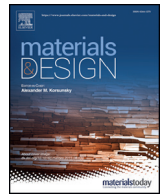

EXPRESS ARTICLE

# An analytical solution for the correct determination of crack lengths *via* cantilever stiffness

Markus Alfreider [a,*], Stefan Kolitsch [b,c], Stefan Wurster [b], Daniel Kiener [a]

[a] *Department Materials Science, Chair of Materials Physics, Montanuniversität Leoben, Leoben, Austria*
[b] *Erich-Schmid Institute of Materials Science, Austrian Academy of Sciences, Leoben, Austria*
[c] *Materials Center Leoben Forschung GmbH, Roseggerstraße 17, Leoben, Austria*

A B S T R A C T

The present work provides an analytic solution for the stiffness to crack length relation in microscopic cantilever shaped fracture specimens based on classical beam theory and substitution of the crack by a virtual rotational spring element. The resulting compact relationship allows for accounting of the actual beam geometry and agrees very well with accompanying finite element simulations. Compared with the only other model present in literature the proposed relationship reduces the deviation between model and data to a maximum of 1.6% from the previous minimum of 15%. Thus, the novel solution will help to reduce the necessity for individual simulations and aim to increase the comparability of elastic-plastic microcantilever fracture experiments in the future.

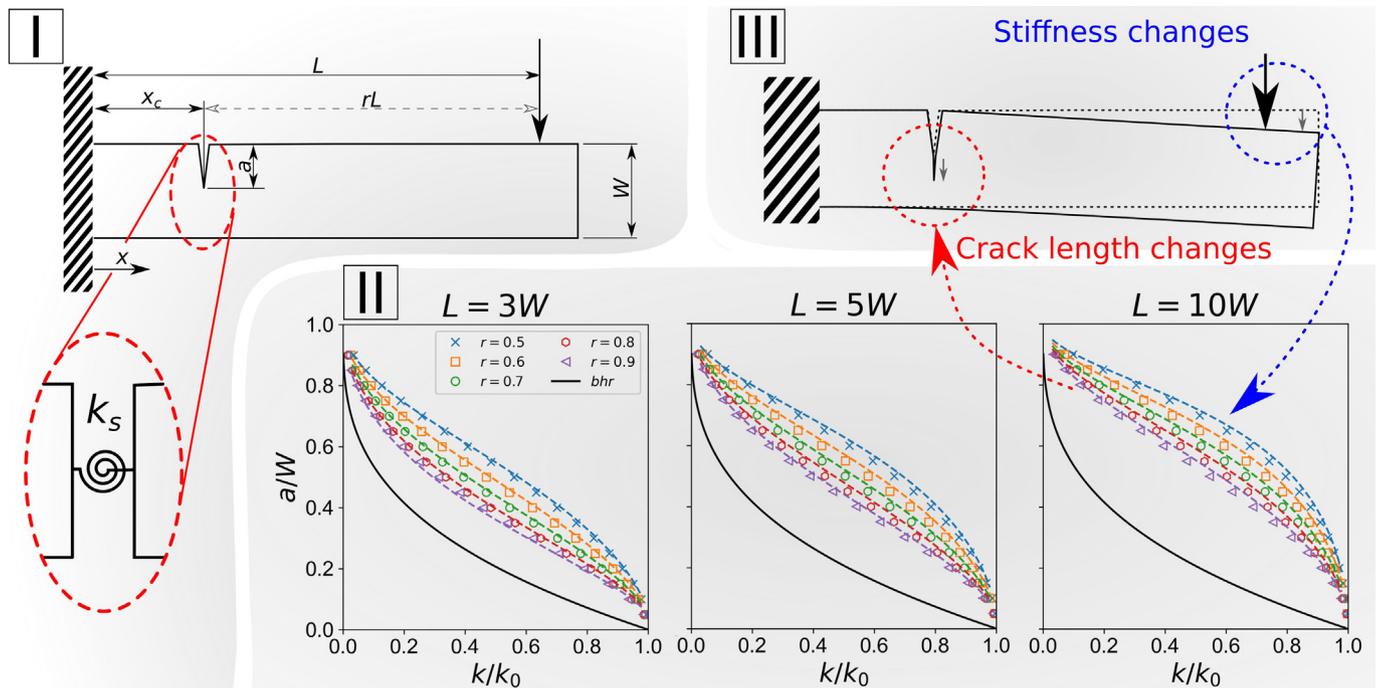

* Corresponding author.
  *E-mail address:* markus.alfreider@unileoben.ac.at (M. Alfreider).

Microscale fracture testing is a rising field in materials science as it enables investigation of previously inaccessible features. A great number of different miniaturized methods have been reported in the literature, from pillar splitting to double cantilever wedging or three-point bending approaches. However, the most prominent realization is the fracture testing of a single notched cantilever geometry [1]. The experimental approach and evaluation is reasonably well understood and agreed on for the case of brittle materials, e.g. hard coatings or ceramics, where for a sufficiently sharp initial notch the concept of linear elastic fracture mechanics holds true. Contrarily, the area of elastic-plastic fracture mechanics is still being explored on the microscopic scale, with a lesser degree of best practices in terms of test setup and data analysis achieved to date. Once the regime of linear elasticity does not hold anymore and noticeable amounts of plasticity take place in the fracture process, elastic-plastic fracture concepts must be involved for analysis. Thereby, independent of the exact experimental testing geometry and analysis method, i.e. crack-tip opening displacement [2] or J-integral [3], the key challenge is always the determination of crack extension.

Most approaches quantify crack extension in an indirect way, through either sequential unloading segments [4] or by an overlaid sinusoidal signal to the applied force [5,6] to measure the change in specimen stiffness. Thereafter, this change in stiffness is used to derive the crack extension. While this seems a trivial elasticity problem and solutions for other geometries are already present in literature [7], for cantilever shaped specimens various different ways were suggested so far. Wurster et al. [4] first assumed a classical Euler-Bernoulli beam with the height being reduced due to the crack extension to describe the stiffness to crack length relation in their experiments. This initial beam height reduction (bhr) approach (shown in detail as Supplementary A) gives a straightforward mathematical formulation. However, comparing it to recent results from finite element modelling and in-situ experiments it appears to deviate rather distinct from the actual relation between stiffness and crack length, as shown in [6]. The reason for this characteristic is because this analysis results in a globally reduced bending stiffness, whereas the stiffness reduction originating from a sharp crack is of local nature and therefore less pronounced.

Ast et al. [5] later employed finite element modelling for their specific geometry, while Alfreider et al. [6] used a polynomial fit through a wide range of finite element data, validated by experiments on various different materials to ensure a certain degree of geometrical and material independence of their approach. However, the correct physical fundamental translation from experimentally determined stiffness changes to actual crack length is still unknown, therefore giving rise to ambiguity in evaluation of nominally analogous experiments in literature.

To model the realistic situation, a concept in recent works by Biondi and Caddemi [8] as well as Alijani et al. [9] is adopted, where such singularities are addressed analytically through Dirac's delta function $\delta(x)$ as a bending slope discontinuity at the crack position by substitution with a virtual rotational spring $k_s$, in a two-dimensional Euler-Bernoulli framework, as shown in feature I of the graphical abstract.

The detailed mathematical derivation of the problem can be found as supplementary material (Supplementary B), but the final compact relation states:

$$\int_0^a \frac{a}{W} Y\left(\frac{a}{W}\right)^2 da = \frac{(k_0/k-1)L}{18\pi(1-\nu^2)r^2} \quad (1)$$

where $a$ is the crack length, $W$ and $L$ are geometric parameters shown in the graphical abstract (feature I), $k$ and $k_0$ are the stiffnesses of the cracked and unnotched beam respectively, $\nu$ is Poisson's ratio, $r = (L - x_c)/L$ (with $x_c$ being the offset of the crack from the base) and $Y(a/W)$ is a geometry factor. This factor has previously been calculated for the shown cantilever geometry by various groups, with only slight deviations among each other as shown by Brinckmann et al. [10]. The first term of Eq. (1) cannot be solved analytically in the general case. However, with nowadays computational efficiency it is trivial to compute the integral approximately, e.g. trapezoidal rule, for a sensible range of $a/W$ and find the corresponding $a$ by interpolation.

To study the accuracy of the model, it was compared with two-dimensional linear elastic finite element simulations. They were conducted using 4-node plane-stress and plane-strain elements and an isotropic material behaviour with an elastic modulus $E_0 = 130$ GPa and a Poisson's ratio $\nu = 0.34$. The cantilever base was considered rigid, with a displacement equal to zero, in accordance with the analytical assumptions taken herein. The calculations were conducted for three different cantilever lengths $3W$, $5W$ and $10W$ with $W = 2$ μm, while $r$ ranges from 0.5 to 0.9 in increments of 0.1, and $a/W$ spans from 0.05 to 0.9 in 0.05 increments. Thus, a total of 540 different simulations were performed. As expected, no difference in $a/W$ over $k/k_0$ data was found in comparison between plane stress and plane strain state, respectively. Hence, all the results summarized in feature II of the graphical abstract are shown in plane strain condition. There, the finite element data is depicted by symbols, and the analytical model is shown by the dotted curves in colours corresponding to the given geometries. The continuous black curve depicts the bhr model [4]. As shown in Supplementary A the bhr model is independent of the cantilever geometry when considered in a normalized manner.

It is evident from the presented data that the proposed analytical model is in very good agreement with the finite element data and the changes in geometry are reflected quite well. To estimate the differences between analytical model and finite element data, the root mean square deviation was calculated for all combinations of $r$ and $L$, revealing the highest deviation to be 1.6% for $r = 0.5$, $L = 10W$. In comparison, the bhr model would deviate by 15% from the data for $L = 3W$, $r = 0.9$, which represents the minimum discrepancy between finite element data and bhr model. Notably, isotropic elasticity was used for convenience. However, due to normalization by the unnotched beam configuration Eq. (1) is independent of elastic properties and therefore, errors originating from elastic anisotropy can be neglected in the given form. In conclusion, the proposed simple and straight forward analytical model describes the observed physical behaviour very well and is recommended to address the stiffness to crack length conversion in the analysis of miniaturized elastic-plastic fracture experiments as schematically depicted in feature III of the graphical abstract.

**CRediT authorship contribution statement**

**Markus Alfreider:** Conceptualization, Formal analysis, Writing - original draft. **Stefan Kolitsch:** Software, Writing - review & editing. **Stefan Wurster:** Methodology, Writing - review & editing. **Daniel Kiener:** Supervision, Writing - review & editing.

**Declaration of Competing Interest**

The authors declare that they have no known competing financial interests or personal relationships that could have appeared to influence the work reported in this paper.

**Acknowledgements**

The authors gratefully acknowledge the financial support under the scope of the COMET program within the K2 Center "Integrated Computational Material, Process and Product Engineering, IC-MPPE" (Projects A2.12 and A1.24). This program is supported by the Austrian Federal Ministries for Transport, Innovation and Technology (BMVIT) and for Digital and Economic Affairs (BMDW), represented by the Österreichische Forschungsförderungsgesellschaft (Funder ID: 10.13039/501100004955), and the federal states of Styria, Upper Austria, and Tyrol. This project has received funding from the European Research Council (ERC) under the European Union's Horizon 2020 - Research and Innovation Framework Programme (Grant No. 771146 TOUGHIT and No. 757333 SpdTuM).

**Data availability statement**

The raw/processed data required to reproduce these findings cannot be shared at this time as the data also forms part of an ongoing study.

**Appendix A. Supplementary data**

Supplementary data to this article can be found online at https://doi.org/10.1016/j.matdes.2020.108914.